\documentclass[12pt,preprint]{aastex}

  \shorttitle{Lorentz Invariance}
  \shortauthors{Boggs et al.}

\begin{document}

  \title{Testing Lorentz Invariance with GRB021206}

  \author{Steven E. Boggs\altaffilmark{1}, C. B. Wunderer, K. Hurley and W. Coburn}

  \affil{
  Space Sciences Laboratory, University of California, Berkeley, CA 94720-7450
  }

  \email{boggs@ssl.berkeley.edu}

  \altaffiltext{1}{Department of Physics, University of California, Berkeley.}

  %\date{\today}

  \begin{abstract}
  Since the discovery of the cosmological origin
  of GRBs there has been growing interest in using these transient events to probe the
  quantum gravity
  energy scale in the range 10$^{16}$--10$^{19}$ GeV, up to the Planck mass scale.
  This energy scale can manifest itself through a measurable modification in the
  electromagnetic radiation dispersion relation for high energy photons originating from
  cosmological distances.
  We have used data from the gamma-ray burst (GRB) of 2002 December 6 (GRB021206) to
  place an
  upper bound on the energy dispersion of the speed of light. 
  The limit on the first-order quantum gravity effects
  derived from this single GRB indicate that the energy scale is in excess of
  1.8$\times$10$^{17}$ GeV. We discuss a program to further constrain the energy scale
  by systematically studying such GRBs.
  \end{abstract}


  %\pacs{98.70.Rz, 03.30.+p, 04.60.-m}

  %\maketitle

  \section{Introduction}

  The general quantum-gravity picture of the vacuum is one of a gravitational medium
  containing
  microscopic quantum fluctuations on size scales comparable to the Planck length,
  $(\hbar G/c^3)^{1/2}=1.6 \times 10^{-33} \rm cm$.
  A number of approaches to quantum gravity
  (noncommutative geometry, loop quantum gravity)
  have independently been demonstrated to modify the
  electromagnetic dispersion relation \citep{ame98,ame03},
  suggesting that first- or second-order spontaneous violation
  of Lorentz invariance at high photon energies might be a general signature of
  quantum gravity phenomenology \citep{sar02}. The effects of this dispersion (reduced
  propagation speeds at high energies) are expected to be very small, unless the signals
  travel over very large distances, and the photon energies are very different from one
  another. 
  The magnitude of this {\it in vacuo} dispersion is set by an assumed energy scale,
  E$_{QG}$, which
  characterizes the size scale of quantum gravitational effects:
  \begin{eqnarray}
  v = \frac{\partial E}{\partial p} \simeq c(1-\xi \frac{E}{E_{QG}} - \mathcal{O}
  (\frac{E}{E_{QG}})^2) ,
  \end{eqnarray}
  where $\xi = \pm1$, but is commonly assumed to be positive \citep{ame02}.
  E$_{QG}$ is generally assumed
  to be on the order of the Planck mass (E$_P \sim$10$^{19}$ GeV); however, theoretical
  work has suggested that this energy scale can be as low as 10$^{16}$ GeV \citep{wit96},
  or even as low as 10$^{3}$ GeV \citep{ark99}. (Note, however, that Lorentz
  invariance was
  preserved in both of these models.)
  With the discovery that GRBs are at cosmological distances \citep{van97},
  it was identified that GRBs could be sensitive to effective energy scales as high as
  the
  Planck mass \citep{ame98}. GRBs can combine high energy photons, millisecond
  time variability, and very large source distances, making it possible to search
  for time delays in GRB lightcurves as a function of energy.

  %A number of quantum gravity scenarios predict an energy dispersion of the speed of
  %light in vacuum due to microscopic quantum gravity fluctuations. This dispersion
  %would result in a time
  %delay of high energy radiation originating from cosmological distances.
  %With the discovery of the cosmological origin of at least some of the
  %gamma-ray bursters (GRBs) \citep{van97}, it was realized that energy dispersions for
  %MeV photons -- measured through millisecond time delays in variability structures
  %in the GRB lightcurve -- could be sensitive to effective energy scales as high as the
  %Planck mass \citep{ame98}.

  The dispersion relation in Eqn. 1 leads to a first-order differential
  time delay for signals of energy $E$ traveling from a source at cosmological distance
  $z$ given by \citep{ell03}:
  \begin{eqnarray}
  \frac{\partial t}{\partial E} \simeq \frac{1}{H_o E_{QG}} \int^{z}_{0}
  \frac{dz}{h(z)} ,
  \end{eqnarray}
  where $t$ is the photon arrival time,
  %$\xi$ is a positive constant of order unity
  %which depends on the specific model,
  \begin{eqnarray}
  h(z) \equiv \sqrt{\Omega _{\Lambda} + \Omega _{M} (1+z)^3},
  \end{eqnarray}
  and $\Omega _{\Lambda}= 0.71$, $\Omega _{M} = 0.29$, $H_o = 72~km~s^{-1}~{Mpc}^{-1}$
  are
  the current best estimates of the
  cosmological parameters \citep{spe03}.
  In some quantum gravity models, the first-order differential time delays vanish,
  and a second-order delay in E$_{QG}$ remains. In this
  case, we would find \citep{ell03}:
  \begin{eqnarray}
  \frac{\partial t}{\partial E} \simeq \frac{2E}{H_o E_{QG}^{2}} \int^{z}_{0}
  \frac{(1+z)dz}{h(z)} .
  \end{eqnarray}

  These time delays hold for any astrophysical source, not just GRBs,
  so several high energy sources exhibiting
  time variability have been used to set a lower limit on E$_{QG}$
  for first-order corrections to the dispersion.
  Pulsed emission from the Crab Pulsar in the
  GeV photon range has been used to set a lower limit of
  E$_{QG}$ $>$ 1.8$\times$10$^{15}$ GeV \citep{kaa99}.
  Initial analysis of GRB timing in the MeV photon range for bursts at
  known redshifts set a lower limit of 10$^{15}$ GeV \citep{ell00}, while a more detailed
  wavelet analysis extended this limit to 6.9$\times$10$^{15}$ GeV \citep{ell03}.
  TeV observations of flares
  in the active galactic nucleus Mkn~421 increased this limit to
  6$\times$10$^{16}$ GeV  \citep{bil99}.
  The current limit using this method is set at 8.3$\times$10$^{16}$ GeV by
  observations of GRB930131 \citep{sch99}
  but the lack of a distance
  measurement makes this subject to considerable uncertainty.
  Other astrophysical methods have placed more stringent constraints on E$_{QG}$
  assuming that the electron dispersion relation is modified as well.
  For example, observations of TeV $\gamma$-rays emitted by blazars
  place a limit on E$_{QG}$ $>$
  3.4$\times$10$^{18}$ GeV by constraining the decay of photons into
  electron-positron pairs \citep{ste03}. Also, the discovery of polarized
  $\gamma$-ray emission \citep{cob03} from the same GRB discussed in this paper
  led to limits of E$_{QG}$ $>$ 10$^{33}$ GeV from birefringence
  constraints \citep{jac03,mit03}. However, it remains possible from the models
  that Lorentz invariance is conserved by electrons and not by photons \citep{ell03b},
  in which case the photon dispersion relation remains a key constraint on
  E$_{QG}$.

  Here we report on the limits set on E$_{QG}$ from GRB021206, an especially bright
  burst with a hard spectrum extending well into the MeV range. While the
  time profile for GRB021206 was quite complex below 2\,MeV, at higher energies it
  exhibited a single, fast flare of photons extending to energies above 10 MeV
  with a duration of $\simeq$15\,ms. The broad
  spectral range measured and the relatively short duration of this flare
  allow us to constrain the lower limit on E$_{QG}$ 
  which is slightly higher than the previous limit
  using this method \citep{sch99}, but consistent with it, considering the uncertainties
  in both cases. 
  It is also comparable to the lower limit set by absorption methods \citep{ste03}.

  \section{Observations}

  We used the Reuven Ramaty High Energy Solar Spectroscopic Imager (RHESSI)
  \citep{lin03} 
  to make these $\gamma$-ray observations of GRB021206. RHESSI has an array of nine
  large 
  volume (300 cm$^3$ each) coaxial germanium detectors with high spectral resolution,
  designed 
  to study solar X-ray and $\gamma$-ray emission (3 keV -- 17 MeV). RHESSI has high
  angular 
  resolution (2$\,^{\prime\prime}$) in the 1$^\circ$ field of view of its optics;
  however, the focal plane detectors 
  are unshielded and open to the whole sky. Thus, while the chances are small that
  RHESSI will 
  see a GRB in its imaging field of view, it measures them frequently in the focal plane 
  detectors themselves, providing the energy and 1-$\mu$s timing of each measured photon.

  Prompt $\gamma$-ray emission from GRB021206 was detected with RHESSI on 2002 
  December 6.951 UT (Fig. 1). This GRB was also observed \citep{hur02a} with the
  Interplanetary 
  Network (IPN).
  Refined measurements by the \it Ulysses \rm and RHESSI spacecraft indicate that
  it had a 25-100 keV fluence of 4.8$\times$10$^{-4}$ erg cm$^{-2}$, making this an
  extremely bright GRB. The IPN localized \citep{hur02b,hur03} 
  GRB021206 to a 8.6 square arcminute error ellipse located 18$^\circ$ from the Sun.
  This solar proximity precluded optical afterglow observations at the time of the
  GRB; however, a candidate radio source was located using the VLA \citep{fra03}.
  Follow up
  optical observations have yet to measure the redshift of the GRB host galaxy,
  which is quite faint.  However, we
  can estimate the redshift (the ``pseudo-redshift'') from the GRB spectral and
  temporal properties
  alone \citep{att03}.  We estimate the spectral parameters of the Band model
  \citep{ban93} to be 
  $\alpha=-1.16$, $\beta=-2.53$, and $E_{break}=648 keV$ and the burst duration to be
  20 s.  This gives $z \simeq 0.3$ . We caution the reader that the redshift
  uncertainty with this novel method could be as high as a factor of 2, which
  would produce a comparable factor of 2
  shift in our first-order lower limit on E$_{QG}$, and a factor of $\sqrt{2}$ shift
  on our
  second order limit.

  Fig. 1 shows the GRB lightcurve divided into 3 energy bins from 0.2--17\,MeV.
  The fast flare seen so clearly
  in the lightcurves above 3\,MeV begins to mix with lower-energy flares below 3\,MeV.
  The 1--2\,MeV range is the lowest energy band where this flare is resolved, though
  at these
  energies it is surrounded by a number of neighboring peaks.
  We cannot rule out unresolved flares $<$2\,MeV contributing to this peak.
  Below 1\,MeV, this feature is
  completely lost in the noise of the other low energy flares comprising the complicated
  lightcurve. Therefore, we focus this analysis on the 1--17 MeV energy range.
  In Fig. 2 we present the lightcurve divided into finer energy bands, and with finer
  temporal resolution. Note that while the number of flare counts in the 7-10 MeV and 10-17 MeV bands is
  significantly smaller than at lower energies, the combined significance of the
  7-17 MeV flare is large, with a chance of 2.6$\times$10$^{-5}$ of being a random Poisson
  fluctuation in the background rate. (i.e. We would expect to randomly see 
  this many peak counts in a single 7.8125 ms time bin about once every 8.7 hours of
  RHESSI background data.)

  For each of
  the energy bands shown in Fig. 2, we analyzed the peaks with two separate methods to
  determine
  the peaking time of the flare in each band. The first method was to bin the event
  data into
  the histogrammed light curves shown in Fig. 2, and then fit the flare to a gaussian
  profile
  in order to characterize the peaking time and the uncertainty. For this analysis we
  chose 7.8125-ms
  wide bins, which are narrow enough to resolve the flare in the 7--10 and 10--17~MeV
  energy ranges.
  The second method we used to determine the peak times was to use 
  the event data directly, and to determine the average (peak) time and standard
  deviation
  for all events in a 50~ms time window centered on the short flare. The results of this
  analysis were relatively insensitive to variations in the window size and center as
  long as the flare dominates the total counts in the window. These two methods
  placed comparable limits on the dispersion, and in Fig. 3 we show the results from
  averaging the
  results of these two separate analysis techniques.

  %From the 1--17\,MeV data, we can determine the best-fit
  %dispersion of the photon arrival times,
  %\begin{eqnarray}
  %\frac{\Delta t}{\Delta E} = 0.40 \pm 0.20~s~GeV^{-1}~~(1-17~MeV).
  %\end{eqnarray}
  %While this fit is consistent with zero time delays at only a 5\% confidence level, we
  %consider this measurement to be an upper limit on the actual dispersion.

  From the peaking times plotted in Fig. 3, we can see that the measured slope would be
  strongly affected by the 1--2\,MeV and 2--3\,MeV data points, and we can not
  preclude the possibility that an additional unresolved flare at
  energies $<$3\,MeV is biasing these two points to earlier peaking times.
  Therefore, we performed a fit of the dispersion for just the
  data $>$3\,MeV. For the 3--17\,MeV band, the time drift of the peak
  is measured to be $\Delta t = 0.0 \pm 4.8$ ms, yielding:
  \begin{eqnarray}
  \frac{\Delta t}{\Delta E} = 0.00 \pm 0.34~s~GeV^{-1}~~(3-17~MeV).
  \end{eqnarray}
  This fit is consistent with a 95\% confidence upper limit on the dispersion of
  \begin{eqnarray}
  \frac{\Delta t}{\Delta E} < 0.7~s~GeV^{-1}.
  \end{eqnarray}
  If we include the 1--3\,MeV data in this fit, the upper limit remains comparable
  at $\frac{\Delta t}{\Delta E} < 0.8~s~GeV^{-1}$.

  Given our upper limit on the time dispersion in Eqn. 6 and the estimated source
  redshift,
  we can calculate the limit on
  E$_{QG}$  for first-order dispersion effects from Eqn. 2.
  This yields a lower limit of E$_{QG}$ $>$ 1.8$\times$10$^{17}$ GeV.
  For second-order dispersion effects from Eqn. 4,
  we can set a lower limit of E$_{QG}$ $>$ 5.5$\times$10$^{7}$ GeV.
  It has been widely speculated that GRBs are detectable to redshifts
  of 10 and beyond \citep{lam00}.
  If the same dispersion were measured for a burst at redshift 10,
  the lower limit would be 33 times higher, or 6$\times$10$^{18}$ Gev,
  which is only slightly smaller than E$_P$.

  \section{Discussion}

  Many GRB energy spectra display hard-to-soft evolution \citep{pre98}.  However,
  this refers to a global trend across the entire GRB time history and across
  the $\sim$25--1000 keV spectrum.  In contrast, our results use the behavior
  of $\sim$ millisecond peaks at energies $>$1000 keV.  In another study \citep{nor00}
  the lag as a function of energy was examined for individual pulses in GRBs.
  A spectral lag was found, characterized by pulses peaking at high energy
  before they peaked at low energy.  However, the pulses in question had
  durations of $\sim$ seconds, the low and high energies were 10's of keV
  and 100's of keV, and the resulting lags had magnitudes of up to several hundred
  milliseconds.  This same study also confirmed an earlier result \citep{fen95},
  namely that pulse widths are narrower at higher energies.  Here, too, however,
  the pulse durations are $\sim$ seconds.  We also note that this
  earlier study, which extended only up to $\sim$1000 keV, made no mention of
  spectral lag \citep{fen95}. The pulses that we are concerned with here
  are orders of magnitude shorter, and orders of magnitude higher
  in energy.  The fact that pulses tend to be narrower with increasing
  energy is an advantage, since our estimate of 
  $\Delta t$ is
  not based on rise times or fall times, but rather on the times of the
  peaks, which are better defined for narrower pulses.  To our knowledge, no studies
  have 
  focussed on such high-energy, short-duration 
  pulses.

  A reliable measurement of E$_{QG}$ will require a systematic study of the
  dispersion as a function of source redshift in order to separate out any residual GRB
  source geometry or emission mechanism effects that can bias the results.
  The ideal instrument to study E$_{QG}$ using GRBs would have coverage to high
  energies ($\geq$10\,MeV), and fine time resolution ($\leq$0.1\,ms). RHESSI,
  designed to study solar flares in the 3\,keV -- 17\,MeV range with 1\,$\mu$s
  photon timing, provides a unique, all-sky GRB monitor for these studies.
  RHESSI nicely complements the HETE-2 and upcoming Swift missions  \citep{ric01,geh00},
  which are able to localize GRBs for follow-up redshift determinations, but
  do not have the spectral range for these studies.
  RHESSI will also provide a low-energy compliment to the upcoming GLAST mission, which
  will also be sensitive for constraining E$_{QG}$ \citep{nor99}.
  We have established a program to study the high energy timing of
  the hundreds of bursts seen in the RHESSI detectors, with a goal of further
  constraining
  E$_{QG}$. The best GRBs for this will have the high energy
  emission as seen in GRB021206 and, ideally, even faster flare peaks.

  \acknowledgments
  The authors are grateful O. Ganor and J. Greiner for useful discussions,
  and to D. Frail for his persistence in localizing this GRB.
  SB and WC are grateful for support under the California Space Institute.
  KH is grateful for Ulysses support under JPL Contract 958056, and for
  IPN support under NASA Grants NAG5-11451, NAG5-12614, and NAG5-13080.

  \clearpage

  \begin{figure}
  \epsscale{.60}
  \plotone{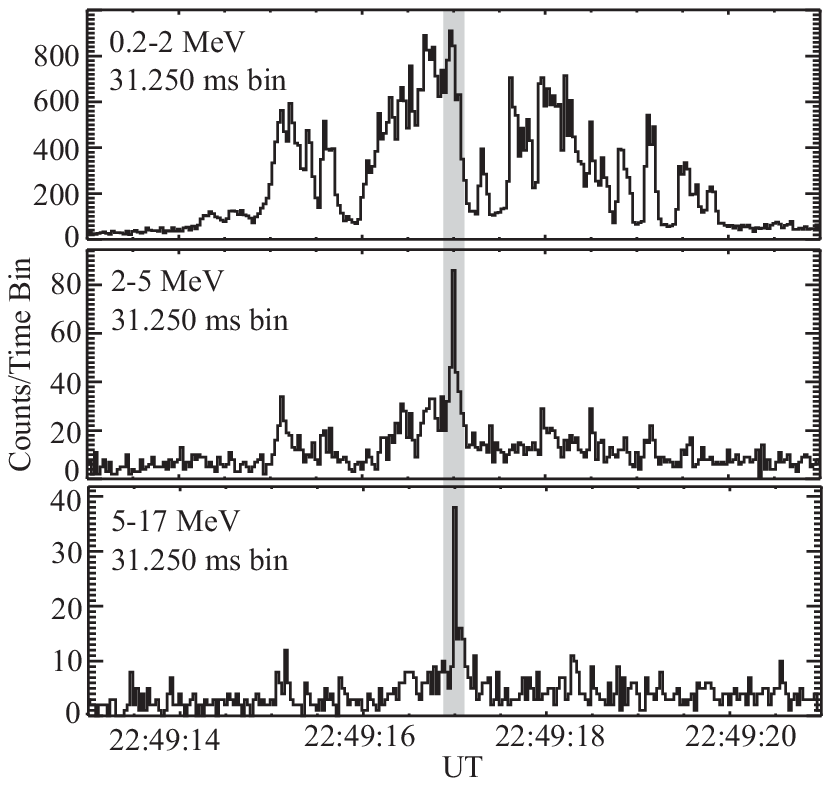}
  \caption{RHESSI lightcurve of GRB021206 in three
   energy bands, spanning 0.2-17\,MeV.\label{fig1}}
  \end{figure}

  \clearpage

  \begin{figure}
  \epsscale{.60}
  \plotone{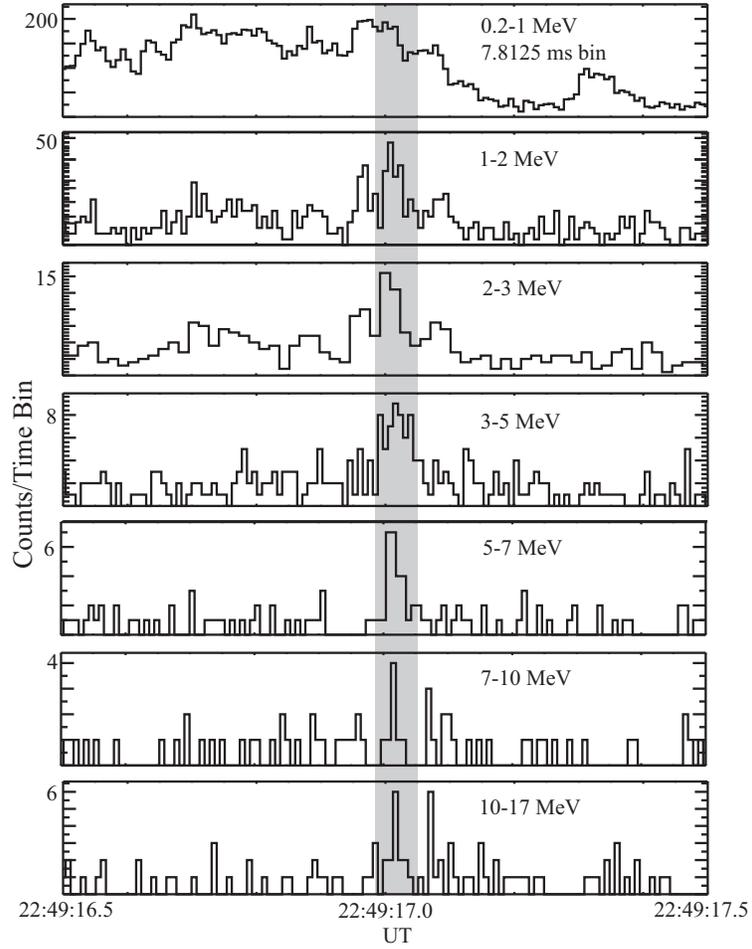}
  \caption{Expanded view of the GRB021206 lightcurve in six energy bands
   spanning 1-17\,MeV at the time of the high energy flare. Time bins of 7.8125\,ms were
   chosen to resolve the flare peak in the highest energy ranges for gaussian
  fitting.\label{fig2}}
  \end{figure}

  \clearpage

  \begin{figure}
  \epsscale{.60}
  \plotone{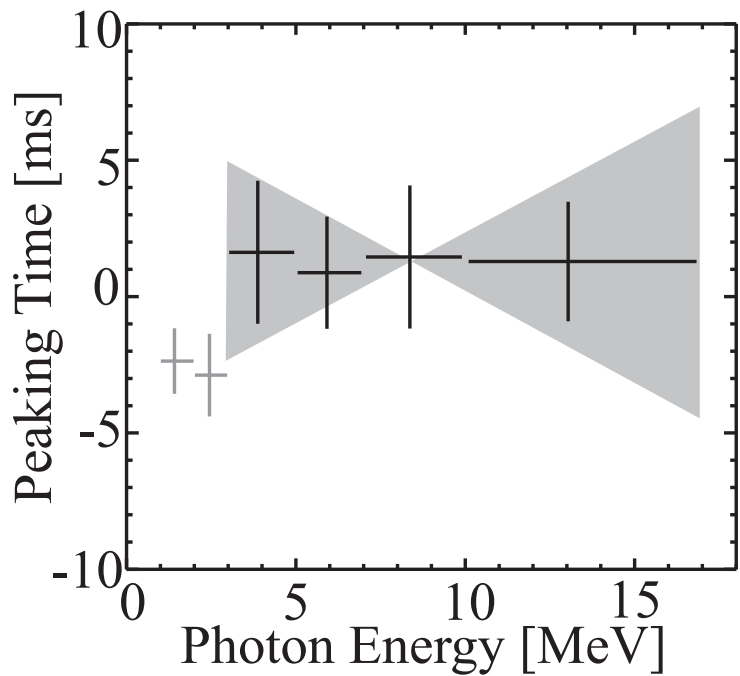}
  \caption{The measured time dispersion of GRB021206
   as a function of energy. The grey region shows the upper and lower limits
   on the possible slope for the 3--17\,MeV data
   ($\pm$95\% confidence). The 1--3\,MeV
   data (shown in grey) were not used in the fit due to potential contamination by
   unresolved flares at these energies (see text).\label{fig3} }
  \end{figure}

  \end{document}